\documentclass[aps,twocolumn,floatfix,preprintnumbers,nofootinbib,superscriptaddress]{revtex4-2}
\usepackage[utf8]{inputenc}  
\usepackage{amsmath}
\usepackage{amssymb} 
\usepackage{multirow}

\usepackage{enumerate}
\usepackage{amsmath}  
\usepackage{tikz}
\usepackage{tipa}
\usepackage{CJKutf8}
\usepackage{feynmf}
\usepackage{slashed}
\usepackage{braket}
\usepackage[toc,page]{appendix}
\usepackage{url}
\usepackage{natbib}
\usepackage{graphicx}
\usepackage[pdftex,bookmarks,linktocpage,pdfpagelabels,plainpages=false,hyperfigures,linkcolor=blue,citecolor=blue]{hyperref} 
\hypersetup{colorlinks=true}

\usepackage{mathrsfs,amssymb}  
\usepackage{cancel}
\usepackage[normalem]{ulem}
\usepackage{array}
\usepackage{booktabs}
\usepackage{verbatim}

\newcommand\barparenb[1]{\overset{%
   \scalebox{0.4}{$(\mkern-1mu-\mkern-1mu)$}}{#1}}

\begin{document}
\begin{flushright}
MI-HET-797~~~~ PITT-PACC-2301
\end{flushright}

\author{P.~S.~Bhupal Dev}
\email{bdev@wustl.edu}
\affiliation{Department of Physics and McDonnell Center for the Space Sciences, Washington University, St.~Louis, MO 63130, USA}

\author{Bhaskar Dutta}
\email{dutta@tamu.edu}
\affiliation{Mitchell Institute for Fundamental Physics and Astronomy, Department of Physics and Astronomy, Texas A\&M University, College Station, TX 77843, USA}

\author{Tao Han}
\email{than@pitt.edu}
\affiliation{PITT PACC, Department of Physics and Astronomy, University of Pittsburgh, Pittsburgh, PA 15260, USA}

\author{Doojin Kim}
\email{doojin.kim@tamu.edu}
\affiliation{Mitchell Institute for Fundamental Physics and Astronomy, Department of Physics and Astronomy, Texas A\&M University, College Station, TX 77843, USA}
\title{Anomalous Tau Neutrino Appearance from Light Mediators in \\
Short-Baseline Neutrino Experiments}

\begin{abstract} 
We point out a new mechanism giving rise to  anomalous tau neutrino appearance at the near detectors of beam-focused neutrino experiments, without extending the neutrino sector. The charged mesons ($\pi^\pm, K^\pm$) produced and focused in the target-horn system can decay to a (neutrino-philic) light mediator via the helicity-unsuppressed three-body decays. If such a mediator carries non-vanishing hadronic couplings, it can also be produced via the bremsstrahlung of the incident proton beam. The subsequent decay of the mediator to a tau neutrino pair results in tau neutrino detection at the near detectors, which is unexpected under the standard three-flavor neutrino oscillation paradigm.  
We argue that the signal flux from the charged meson decays can be significant enough to discover the light mediator signal at the on-axis liquid-argon near detector of the DUNE experiment, due to the focusing of charged mesons. In addition, we show that ICARUS-NuMI, an off-axis near detector of the NuMI beam, as well as DUNE, can observe a handful of tau neutrino events induced by beam-proton bremsstrahlung.
\end{abstract}

\maketitle

\noindent {\bf Introduction.} 
The discovery of neutrino oscillations has proven the non-zero neutrino masses~\cite{Super-Kamiokande:1998kpq,SNO:2001kpb,DoubleChooz:2011ymz,DayaBay:2012fng,RENO:2012mkc,Workman:2022ynf}, requiring an extension of the particle spectrum and/or interactions in the Standard Model (SM). 
Precise measurements of the oscillation phenomena allow us to not only determine physics parameters in the neutrino sector but also obtain hints to beyond-the-Standard-Model (BSM) physics upon the observation of any deviation from the expectations of the standard three-flavor oscillation scheme.
Upcoming neutrino experiments are promising along this line as they are expected to measure the neutrino oscillations more precisely with their high-capability detectors. 

Of them, the appearance of tau-flavor neutrinos is receiving growing attention, especially in the short-baseline neutrino experiments. 
According to the neutrino oscillation theory, the $\nu_\tau$ appearance probability, in particular, with regard to the muon neutrino beam, is given by~\cite{Workman:2022ynf}
\begin{align}
P_{\mu\to\tau}=\sin^2(2\theta_{23})\sin^2 \left[1.27 \frac{\left({\Delta m_{23}^2\over {\rm eV}^2}\right)\ \left({L\over {\rm km}}\right)}{E/{\rm GeV}} \right]\, ,
    \label{eq:prob}
\end{align}
where $\theta_{23}$ and $\Delta m^2_{23}$ are the atmospheric mixing angle and mass-squared splitting  respectively, $L$ is the baseline length and $E$ is the neutrino energy. 
For typical baseline ($L\sim$ 0.5 km) and neutrino energy ($E\sim$ 2 GeV) realized in beam-based neutrino experiments, one can see from Eq.~\eqref{eq:prob} that the chance of observing any $\nu_\tau$-induced events in their near detectors is very small.  
Heavy meson (such as $D$, $D_s$) decays give $\nu_\tau$, but their production rates are negligible in these experiments. 
Therefore, the $\nu_\tau$ appearance at near detectors is {\it anomalous} in itself and is a ``smoking-gun'' signature of new physics. 

In contrast, there are many scenarios for BSM physics that could lead to $\nu_\tau$ in the final state. 
In the simplest extension, the existence of sterile neutrinos mixing with $\nu_\tau$ would alter the probability shown in Eq.~(\ref{eq:prob}), allowing for non-negligible $\nu_\tau$ detection rates at near detectors. 
However, non-conventional oscillation phenomena arising in these scenarios have been constrained by other existing experiments, either using atmospheric neutrinos such as Super-K~\cite{Super-Kamiokande:2014ndf}, IceCube~\cite{IceCube:2017ivd}, and ANTARES~\cite{ANTARES:2018rtf}, or using accelerator-produced neutrinos in far detectors of OPERA~\cite{OPERA:2019kzo}, MINOS~\cite{MINOS:2017cae}, T2K~\cite{T2K:2019efw}, and NO$\nu$A~\cite{NOvA:2021smv}. 
Therefore, the regions of sterile neutrino parameter space to be explored by upcoming beam-based neutrino experiments like DUNE would be limited~\cite{Coloma:2017ptb,DeGouvea:2019kea,Ghoshal:2019pab,Machado:2020yxl,Coloma:2021uhq,MammenAbraham:2022xoc}. 

In this Letter, we point out the novelty of 
light dark-sector scenarios allowing for an anomalous $\nu_\tau$ appearance at the near detectors of beam-based neutrino experiments, without modifying the neutrino sector. In particular, we discuss the $\nu_\tau$ detection prospects at the DUNE near detector~\cite{DUNE:2021tad} and ICARUS-NuMI~\cite{ICARUS:2023gpo} experiments in the context of $\nu_\tau$-philic mediators.\footnote{Similar-origin $\nu_\tau$ events can also arise in higher-energy accelerator experiments~\cite{Kling:2020iar, Batell:2021snh}.}
From an experimental perspective, we address the need for $\nu_\tau$-optimized target-horn configurations, 
from which any new physics scenarios allowing for upscattering processes of new particles ($e.g.$, models of inelastic dark matter) may benefit. 

\medskip

\begin{figure}[t]
    \centering
    \includegraphics[width=8.8cm]{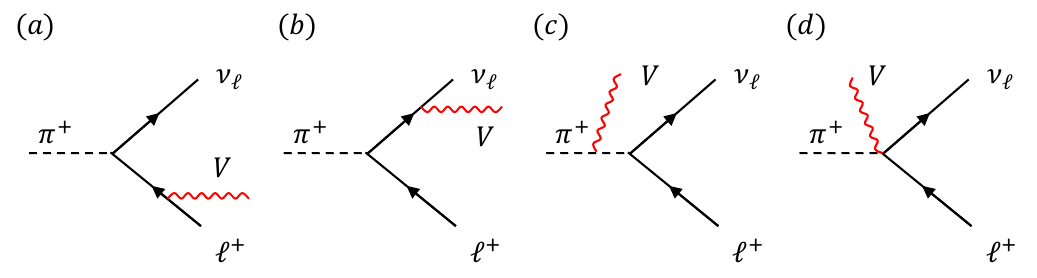}
    \caption{Various three-body decay dynamics of a charged pion (or kaon) to a massive vector boson $V$: $(a)$ emission from the charged lepton, $(b)$ emission from the neutrino, $(c)$ emission from the charged pion, and $(d)$ contact interaction and QCD-originating pion-structure-dependent contributions.}
    \label{fig:diagrams}
\end{figure}

\noindent {\bf Proposal outline.}
The main idea is based upon the recent realization that the exotic three-body decays of charged mesons, such as $\pi^\pm$ and $K^\pm$, can be great sources of dark-sector particles in the beam-based neutrino experiments~\cite{Dutta:2021cip}.
We envision a situation where a new mediator, 
which can be either a (pseudo)scalar or a (massive) vector, is produced from the three-body decay of charged mesons and decays to a $\nu_\tau$ pair, {\it i.e.,} 
\begin{equation}
\pi^\pm/K^\pm \to \ell^{\pm} \barparenb{\nu}_\ell \ V\quad  {\rm with}\ \ V \to \nu_\tau \bar{\nu}_\tau .
\end{equation}
Fig.~\ref{fig:diagrams} depicts various decay dynamics,  with $V$ representing a massive vector boson. 
Such three-body decays of charged mesons are free from the helicity suppression from which their corresponding two-body decays severely suffer, and thus their resultant branching ratios (BRs) can be significantly enhanced despite the extra phase-space suppression in the three-body decays~\cite{Barger:2011mt,Carlson:2012pc,Laha:2013xua,Bakhti:2017jhm,Krnjaic:2019rsv}. 
The BR enhancement can be even larger with a massive (light) vector mediator, owing to the existence of the longitudinal polarization mode giving rise to the terms proportional to $m_{\pi,K}^2/m_V^2$ in the decay width~\cite{Carlson:2012pc}. If $V$ has hadronic couplings, the enhancement can be even more significant; see Appendix for details.

Furthermore, it is important to note that in a beam-focused neutrino experiment, the charged mesons created by the proton collision on a target enter the magnetic horn system and get focused before they decay.
Therefore, the focusing feature in combination with the above-discussed BR enhancement can lead to a substantially enhanced flux of $\nu_\tau$'s at the near detectors, resulting in the surprising emergence of $\nu_\tau$-induced charged-current events. 

\medskip

\noindent {\bf Example scenarios.}
The anomalous $\nu_\tau$ production arises due to the couplings of a new mediator to the SM lepton sector.
While such mediators can be either (pseudo)scalar or vector bosons, and our discussion is generically applicable irrespective of underlying model details, we shall consider models of a massive vector boson $V$ for concreteness. 
Focusing on the flavor-diagonal interactions, we write the relevant interactions as
\begin{equation}
    \mathcal{L}_{\rm int} \supset \sum_f g_V x_f V_\mu \bar{f}\gamma^\mu f\,, 
    \label{eq:xf}
\end{equation}
where $g_V$ is the new coupling parameter associated with the vector mediator $V$ and $x_f$ denotes the gauge charge of SM fermion species $f$. 
Two benchmark scenarios are considered as follows, depending on whether the mediator couples to only leptons or to both quarks and leptons: 

\medskip

\noindent {\it Case (i):~Neutrino-philic model}. Here the mediator only couples to neutrinos at tree level. In this case, $x_f=1$ for $f=\nu_e$, $\nu_\mu$, $\nu_\tau$ and $x_f=0$ otherwise. Examples of such models include $\nu$-philic $U(1)$~\cite{Farzan:2016wym, Farzan:2017xzy, Abdallah:2021npg} and $\nu_R$-philic $Z'$~\cite{Chauhan:2020mgv, Chauhan:2022iuh}. 
Another popular class of models with the $\nu$-philic nature is that the mediator couples to both neutrinos and charged leptons at tree level, but not to quarks. Typical examples are $U(1)_{L_\alpha-L_\beta}$ models~\cite{He:1990pn,He:1991qd,Araki:2012ip}. In this case, the $\nu_\tau$ production rate is enhanced, because the mediator boson can be produced from either the neutrino or the charged-lepton leg in charged-meson decay. However, this scenario does not give better sensitivity, compared to the other cases, because of relatively low-energy $\nu_\tau$'s from electron bremsstrahlung. Therefore, we will not show the results separately for the $U(1)_{L_\alpha-L_\beta}$ scenario. 

\medskip

\noindent
{\it Case (ii):~$B-L$ model}. The mediator can couple to both quarks and leptons, such as in various $U(1)_{X}$ models~\cite{Appelquist:2002mw, Oda:2015gna}, $X=B-L$ being the most popular choice~\cite{Davidson:1978pm, Marshak:1979fm}. Although mesons carry no baryon number, they can couple to the mediator via kinetic mixing (see Appendix). Therefore, the production rate in this scenario is maximal because $V$ can also be emitted from the charged-meson leg, in addition to the final-state leptons, as shown in Fig.~\ref{fig:diagrams}. Although the constraints on such a light-mediator coupling to quarks are rather severe (coming from beam-dump experiments), we can still get a sizable rate at ICARUS-NuMI and/or DUNE with the allowed couplings. Moreover, in some versions of $U(1)_X$ models like in $U(1)_{I_{3R}}$~\cite{Dutta:2019fxn, Dutta:2020enk}, most of the low-energy constraints can be weakened or even evaded altogether. 
Besides the $B-L$ model with a flavor-universal coupling to both quarks and leptons in Eq.~(\ref{eq:xf}), we also consider another interesting incarnation, i.e. the $B-3L_\tau$ model~\cite{Ma:1997nq}, with $x_f=1$ for $f={\rm quarks}$, $x_f =-3$ for $f=\tau, \nu_\tau$, which is best probed via $\nu_\tau$ detection. 

We note that a similar idea appeared in Ref.~\cite{Bakhti:2018avv} in the pure leptophilic $Z^\prime$ context.
As delineated above, our study, however, highlights the generic BR enhancement feature extended to the mediators carrying hadronic couplings and hence improved $\nu_\tau$ discovery potentials. 
Furthermore, we shall show the dependence of the $\nu_\tau$ appearance on the target-horn configuration, motivating the $\nu_\tau$-optimized mode of DUNE to exploit the physics opportunity of this sort. 

\medskip

\noindent {\bf Benchmark study.}
To investigate the detection prospects of $\nu_\tau$-induced events, we consider two beam-focused neutrino experiments, DUNE and ICARUS-NuMI, as concrete examples. 
In both experiments, a 120-GeV proton beam strikes a graphite target, and the charged mesons produced by proton collisions are focused in their magnetic horn systems. 
The focused charged mesons are allowed to decay to neutrinos within 204 meters for DUNE~\cite{DUNE:2020lwj} and 715 meters for ICARUS-NuMI~\cite{Adamson:2015dkw}, before reaching the shielding and rock area. 
The DUNE near detector complex consists of three components out of which we consider the liquid-argon near detector (ND-LAr) that adopts the liquid argon time projection chamber technology which is also adopted for the ICARUS-NuMI detector.  
The DUNE detector is located 575 meters away from the DUNE target on the beam axis~\cite{DUNE:2020lwj}, while the ICARUS-NuMI detector is located $800$ meters away from the NuMI target at an angle of $\sim 6^\circ$ from the NuMI beam line. 
We summarize some key specifications of both experiments and their detectors in Table~\ref{tab:detectorspec}. 

\begin{table}[t]
    \centering
    \begin{tabular}{c|c|c}
    \hline \hline
         & DUNE~\cite{DUNE:2020lwj} & ICARUS-NuMI~\cite{Antonello:2013ypa,Adamson:2015dkw}  \\
    \hline
    Beam energy & 120 GeV & 120 GeV\\
    Dist. to dump  & 204 m & 715 m \\
    Dist. to detector & 575 m & 800 m\\
    Detector angle & On axis & $\sim 6^\circ$ \\
    Active volume & \multirow{2}{*}{$3\times 4 \times 5$} & $2.96\times 3.2 \times 18$ \\
    $(w\times h \times l)~[{\rm m}^3]$ & & ($\times$ 2 modules)\\
    \hline \hline
    \end{tabular}
    \caption{Key specifications of the DUNE (ND-LAr) and ICARUS-NuMI experiments and their detectors. The two distances are measured from the beam target, and the detector angle is measured from the beam axis.}
    \label{tab:detectorspec}
\end{table}

\begin{figure*}[ht!]
    \centering
    \includegraphics[width=0.45\textwidth]{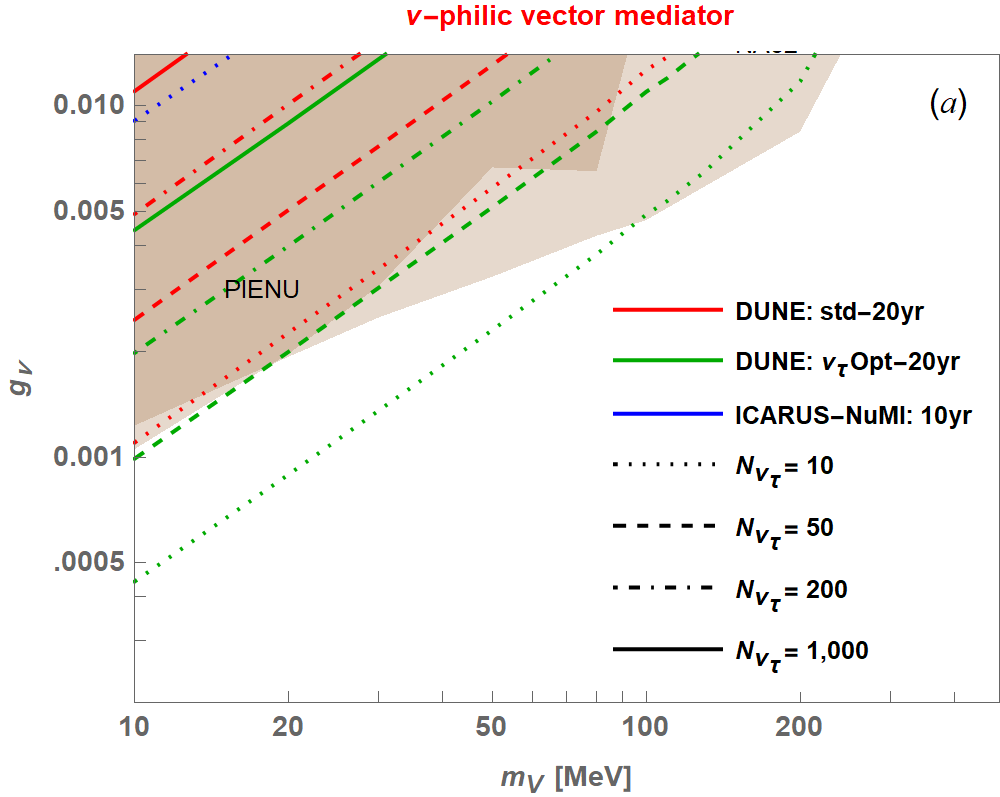} 
    \includegraphics[width=0.45\textwidth]{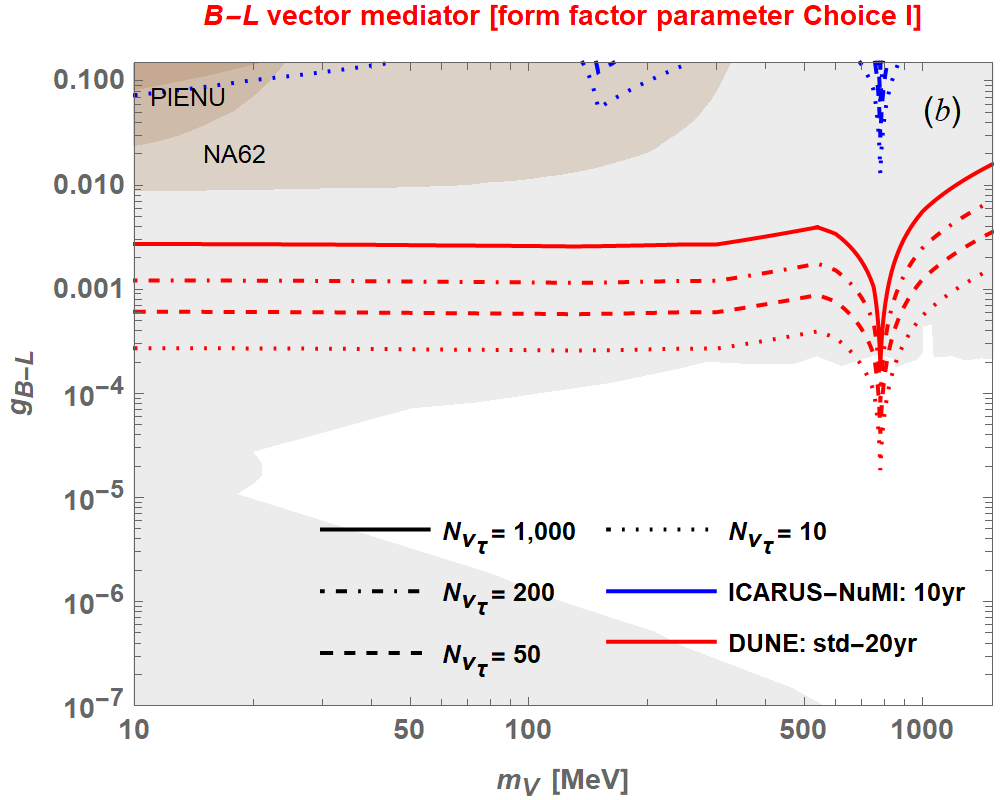} \\
    \includegraphics[width=0.45\textwidth]{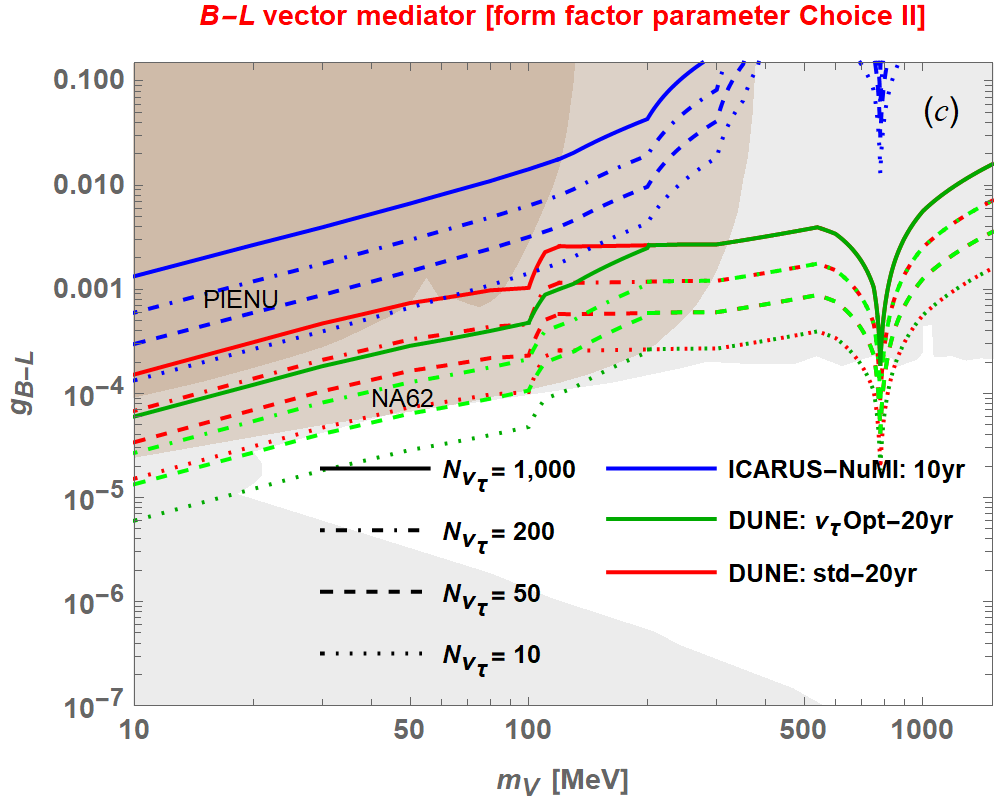}
    \includegraphics[width=0.45\textwidth]{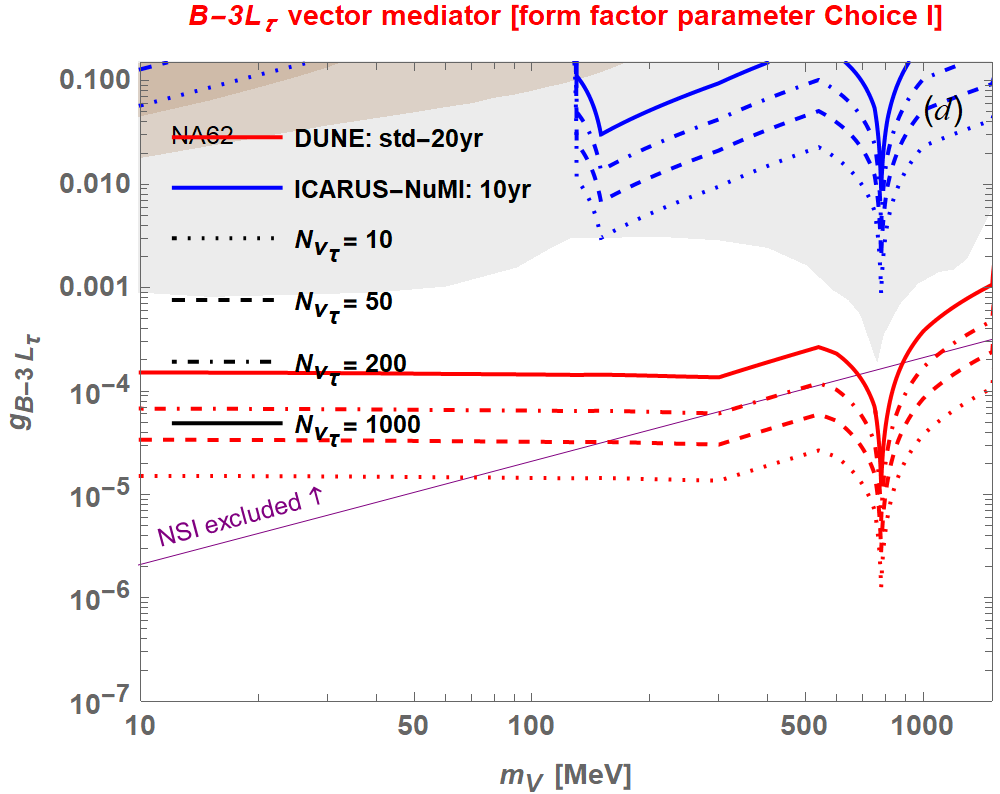} 
    \caption{Sensitivities of DUNE-standard-mode (red), DUNE-$\nu_\tau$-optimized-mode (green), and ICARUS-NuMI (blue) to $\nu_\tau$ events expected in ($a$) the $\nu$-philic  model, ($b$) $B-L$ model with charged meson form factor parameter Choice I, ($c$) $B-L$ model with charged meson form factor parameter Choice II, and ($d$) $B-3L_\tau$ model with charged meson form factor parameter Choice I.  See the text and Appendix for our form factor parameter choices. 
    The gray-shaded regions in the $B-3L_\tau$ and $B-L$ models show the existing limits coming from either beam-dump, or accelerator/reactor-based neutrino scattering experiments, compiled in {\it e.g.}, Refs.~\cite{Ilten:2018crw, Bauer:2018onh, Heeck:2018nzc, Kling:2020iar}. 
    Additional limits (brown-shaded) come from the search for three-body pion decays $\pi^+\to 
    \ell^+ \nu_\ell V$ (PIENU)~\cite{PIENU:2021clt} and three-body kaon decays $K^+ \to \mu^+ \nu_\mu V$ (NA62)~\cite{NA62:2021bji} with $V$ decaying invisibly. The NSI constraint~\cite{Heeck:2018nzc} (purple dashed) is also shown for the $B-3L_\tau$ case, but in principle, it can be evaded by going beyond the minimal model.  
    } 
    \label{fig:B-3Ltau}
\end{figure*}

Tagging a $\nu_\tau$ event requires the incoming $\nu_\tau$ to be energetic enough (with $E_\nu\gtrsim 3.5$ GeV) to upscatter to a tau lepton. 
The DUNE Collaboration is investigating the potential of $\nu_\tau$-optimized target-horn configuration to maximize the focusing of energetic charged mesons that possibly source energetic tau neutrinos~\cite{DUNE:2020lwj}. 
In our analysis, we also consider the $\nu_\tau$-optimized mode of DUNE (in addition to the standard mode) to assess its impact on detection prospects. 

In our simulation study, we first estimate the production rates of charged and neutral mesons ($\pi^\pm$, $\pi^0$, $K^\pm$, and $\eta$~)\footnote{We do not consider the $D$ mesons, since they are estimated to contribute less than one event for our setup.} using the \texttt{GEANT4}~\cite{GEANT4:2002zbu} code package with the \texttt{QGSP\_BIC} physics list; we set a 120 GeV proton beam to impinge on a graphite rod whose specification is similar to that for DUNE. 
The focusing-horn effects are modeled, based on simple assumptions described in Appendix.
In addition to the charged-meson-induced production, we include contributions from neutral mesons ({\it e.g.}, $\pi^0$ and $\eta$) and the proton beam if the dark-sector model of interest allows for hadronic interactions.  
To describe the mediator emission from an incident proton, we follow the formalism in Ref.~\cite{Foroughi-Abari:2021zbm}. 

Once a mediator is produced by the decay of mesons or the proton bremsstrahlung, we assume it to promptly decay to a $\nu_\tau$ pair as per the BR that the model under consideration predicts. 
We then check whether the $\nu_\tau$ enters the detector active volume, and estimate the expected number of $\nu_\tau$-induced events for such a $\nu_\tau$ by multiplying the nucleon number density in liquid argon by the charged-current scattering cross-section for the associated $E_{\nu_\tau}$~\cite{Jeong:2010nt} and the detector length. 

\medskip

\noindent {\bf Results.} As outlined earlier, we present our results in the context of $\nu$-philic and $B-L$  interactions. To make the point of $\nu_\tau$ appearance, we also adopt a $B-3L_\tau$ interaction model. Our results can straightforwardly be extended to other gauged $U(1)$ models, as well as to other light (pseudo)scalar states.

In Fig.~\ref{fig:B-3Ltau}($a$), we show the sensitivities of the DUNE standard mode (red curves) and ICARUS-NuMI (blue curves) to the coupling $g_\nu$ as a function of mediator mass for the $\nu$-philic case. 
The detectors are exposed to $2\times 10^{22}$ POTs for DUNE~\cite{private2} and $10^{22}$ POTs for ICARUS-NuMI.\footnote{$10^{22}$ POTs correspond to a $\sim 10$ year data collection of DUNE. A similar level of POTs is achievable for ICARUS-NuMI~\cite{NuMI}.}
To develop our intuition on the number of signal $\nu_\tau$ events ($N_{\nu_\tau}$) allowed by the model, we show four representative numbers, $N_{\nu_\tau}=10$ (dotted), $N_{\nu_\tau}=50$ (dashed), $N_{\nu_\tau}=200$ (dot-dashed), and $N_{\nu_\tau}=1,000$ (solid) before applying the $\tau$-lepton identification efficiency.  
Therefore, one can assess the expected number of $\nu_\tau$ events by multiplying each of the numbers by the actual identification efficiency. As we will discuss later, the cases of $N_{\nu_\tau}=1,000$ and 200 describe the realistic sensitivity reaches that are achievable with the existing $\tau$ tagging and background rejection efficiencies.
The other cases with lower $N_{\nu_\tau}$ numbers should be understood as projected sensitivities anticipating future improvements on the $\tau$ tagging and background rejection techniques.
The brown-shaded regions show the limits from the search for three-body pion decays $\pi^+\to \ell^+ \nu_\ell V$ (PIENU)~\cite{PIENU:2021clt} and three-body kaon decays $K^+ \to \mu^+ \nu_\mu V$ (NA62)~\cite{NA62:2021bji} with $V$ decaying invisibly.

This result suggests that DUNE operating in the standard mode for 20 years may observe a handful of $\nu_\tau$ events if $m_V\lesssim 50$~MeV. 
As briefly discussed earlier, the DUNE target-horn configuration in the standard mode is not optimized for the observation of tau neutrinos. Therefore, if DUNE allows for the $\nu_\tau$-optimized mode measurement, $\nu_\tau$'s preferentially reaching the detector are energetic enough to upscatter to $\tau$'s, hence more $\nu_\tau$ events would be detected.  
To see this potential, we perform the $\nu_\tau$-optimized simulation according to the simulation scheme that we described earlier and report the expected sensitivity reaches by the green curves in Fig.~\ref{fig:B-3Ltau}($a$). 
For purposes of a fair comparison, the same exposure and $\tau$ identification efficiency are imposed. 
As expected, more $\nu_\tau$ events would be observed, and quantitatively $N_{\nu_\tau}$ would be increased by up to an order of magnitude over the allowed mass range. 

When it comes to the $B-L$ model, hadronic production channels are opened, viz. (i) the proton beam can radiate a $V$ via bremsstrahlung process, (ii) $\pi^\pm$ and $K^\pm$ can internally emit a $V$, and (iii) $\pi^0$ and $\eta$ can decay to $V$ in association with a photon. 
As mentioned earlier, since mesons do not carry non-trivial baryon numbers, production (ii) through hadronic interactions [{\it i.e.},  Fig.~\ref{fig:diagrams}($c,d$)] can arise via the usual kinetic mixing with the SM photon, whereas production (iii) arises through anomaly hence no kinetic mixing suppression~\cite{Bauer:2018onh}. One may argue that the hadronic contribution in (ii) would be subdominant. However, some of the parameters involved in the hadronic interactions in the charged-meson three-body decay are undetermined~\cite{Khodjamirian:2001ga} and the resulting contributions can be sizable, depending on the parameter choices. We present the parameter definitions and technical details in Appendix. We take two reference hadronic form factor parameter ($c_i$'s) choices for illustration purposes:
\begin{eqnarray}
    {\rm I}:&& c_1 = 0.1~{\rm GeV},\, c_2= c_4=10~{\rm GeV}^{-1}, \nonumber \\
    {\rm II}:&& c_1 = 10^2~{\rm GeV},\, c_2=c_4=10^4~{\rm GeV}^{-1}. \nonumber
\end{eqnarray}
By contrast, the signal fluxes from (i) and (iii) are proportional to $g_{B-L}^2$. Therefore, for $m_V\lesssim 500$~MeV, the contribution from (ii) can exist even for small $g_{B-L}$ values suppressing the contributions from (i) and (iii).  
Fig.~\ref{fig:B-3Ltau}($b$) shows the sensitivity reaches with Choice I. Existing limits compiled in {\it e.g.}, Ref.~\cite{Bauer:2018onh} are shown by the gray region and the brown regions show the PIENU~\cite{PIENU:2021clt} and NA62~\cite{NA62:2021bji} limits interpreted with parameter choice I. We find that mesonic contributions are subdominant and proton bremsstrahlung contributions govern the sensitivity reaches, allowing us to explore a certain range of new parameter space without the $\nu_\tau$-optimized mode of DUNE. 
On the other hand, the sensitivity reaches in Fig.~\ref{fig:B-3Ltau}($c$) suggest that the scenario with parameter choice II would allow the charged-meson contributions to dominate over the other contributions. Therefore, the $\nu_\tau$-optimized mode measurement (green curves) would benefit from the charged mesons significantly. 

Finally, Fig.~\ref{fig:B-3Ltau}($d$) shows the sensitivity reaches in the context of the $B-3L_\tau$ model with parameter choice I. Since charged mesons, muons, and electrons are uncharged under $B-3L_\tau$, the $V$ emission from the charged mesons occurs again via kinetic mixing. Therefore, the charged-meson contributions with choice I are negligible. 
The gray region shows the existing limits compiled in Ref.~\cite{Bauer:2018onh}, plus the constraint from the measurement of the tau neutrino rate at DONuT~\cite{Kling:2020iar}. We have also shown the constraint from the neutrino non-standard interactions (NSI)~\cite{Heeck:2018nzc,Coloma:2020gfv} by the purple line. Note that the NSI constraint is somewhat stronger than the other constraints shown in the plot; however, it is model-dependent and can be potentially avoided, {\it e.g.}, in the presence of an additional scalar NSI~\cite{Ge:2018uhz, Babu:2019iml} which could cancel the negative contribution from vector NSI induced by $V$. 
In any case, our results suggest that ICARUS-NuMI would explore a part of the unexplored regions of parameter space. DUNE, even {\it without} the $\nu_\tau$-optimized mode, would extend the sensitivity reach much further, even surpassing the NSI constraints (if applicable) and probing up to $m_V\sim3.5$~GeV, before the $Z\to \tau\tau$ constraints take over.

\medskip

\noindent {\bf Discussion and Conclusion.} Our study is based on a few assumptions at the production and detection of $\nu_\tau$ signals. To estimate the expected $\nu_\tau$ flux, we relied on simple empirical modeling of the focusing effects on the charged mesons and their decays.  
For a more precise estimate of not only the $\nu_\tau$ signal but also other BSM signals ({\it e.g.}, light dark matter), it is highly desired to perform a dedicated simulation where one keeps track of the charged-meson behavior in the focusing horn system. 
Furthermore, the $\nu_\tau$-optimized measurement mode and related simulation study are highly encouraged as it not only improves the sensitivity to the anomalous $\nu_\tau$ signal but also benefits other BSM signals accompanying upscattering processes ({\it e.g.}, inelastic dark matter). 

Regarding the $\nu_\tau$ signal detection, we assumed an ideal $\tau$ identification efficiency. In practice, we will be limited by statistics, since the LAr detectors cannot identify tau neutrinos on an event-by-event basis. Taus are detected through their decay products (either hadronic or leptonic) and any misidentification would cause backgrounds. This has been considered a major challenge in $\nu_\tau$ detection at DUNE in the context of BSM neutrino oscillation studies~\cite{DeGouvea:2019kea,Ghoshal:2019pab,Giarnetti:2020bmf,Coloma:2021uhq}. 

Let us first discuss the hadronic channel.
According to the $\nu_\tau$-detection study in the hadronic channel at DUNE far detectors, $\sim200$ other neutrino events would be mistagged~\cite{DeGouvea:2019kea}. Using the distance and volume of the DUNE near detector, we estimate up to $\sim10^6$ mistagged events, which would therefore require at least a few thousand events in the hadronic channel to have sensitivity to BSM physics. 
However, the recent development of the $\nu_\tau$ identification in the $\rho$ meson channel -- which is based on machine-learning techniques with various kinematic observables -- suggests that a (nearly) vanishing mis-identification rate is achievable while tagging $\sim 5\%$ of signal $\nu_\tau$ events~\cite{private3,TautagAdam}, {\it i.e.}, a zero-background analysis is effectively possible. 
With Br($\tau \to \rho)\approx 26\%$, neutrino experiments can be sensitive to $\sim 180$ $\nu_\tau$ scattering events.  
Along this line, we report the lines of $N_{\nu_\tau}=200$ in Fig.~\ref{fig:B-3Ltau}.

Similar machine-learning-based strategies get through for the leptonic channels. For example of the electron channel, recent studies suggest that requirements enabling a $\sim 1-2\%$ level tagging efficiency would result in (almost) complete background rejection~\cite{private3,SnowmassTalk}. 
Therefore, considering Br$(\tau \to e)\approx 18\%$, we see that about $1,000$ $\nu_\tau$ CC events are required to have 90\% CL sensitivity.
In this context, the lines of $N_{\nu_\tau}{=1,000}$ in Fig.~\ref{fig:B-3Ltau} can therefore be interpreted as the sensitivity reaches that are achievable with the existing $\tau$ identification techniques in the electron channel.

While existing techniques require a couple of hundreds to a thousand signal events to have experimental sensitivity, we emphasize that there are various handles to improve the $\tau$ identification techniques. The level of hadronic activities induced by $\nu_\tau$ scattering vs. $\nu_{\mu/e}$ scattering could be a good discriminator. 
The DUNE Collaboration is carefully investigating $\nu_\tau$-identification and related backgrounds in all three major $\tau$-decay channels ({\it i.e.}, $\rho$-meson, muon, and electron decay channels), in combination with more sophisticated machine-learning techniques and additional event information from the SAND detector in the ND complex~\cite{DUNE:2020ypp,Kosc:2021huh}. 
Moreover, both the $\tau$ tagging and background rejection efficiencies perform better with increasing energy, so the measurements in the $\nu_\tau$-optimized mode will benefit from further improved identification-vs.-mistagging efficiencies, requiring fewer numbers of signal events to reach the same level of sensitivity limits. 
Along this line, our projected optimistic) anticipations are displayed in Fig.~\ref{fig:B-3Ltau} by the other sensitivity lines with $N_{\nu_\tau}=50$ and 10 that can be respectively attained by $\sim8\%$ and $\sim40\%$ tagging efficiencies in all three major channels together with a nearly zero-background environment. 

In conclusion, we expect that a more precise assessment of the LAr detection prospects of anomalous $\nu_\tau$ events will be available in the near future, and this letter provides new physics motivations to do so.

\medskip

\noindent {\bf Acknowledgments.} We thank Antoni Aduszkiewicz, Brian Batell, Pilar Coloma, Matheus Hostert, Yu Seon Jeong, Kevin Kelly, Felix Kling, Pedro Machado,  Alexandre Sousa, and Adrian Thompson for useful discussions. PSBD, BD, and DK acknowledge the warm hospitality provided at PITT PACC, where part of this work was done. 
The work of PSBD is supported in part by the U.S.~Department of Energy under Grant No.~DE-SC0017987 and by a URA VSP fellowship.
The work of BD and DK is supported by the U.S.~Department of Energy Grant DE-SC0010813. 
The work of TH is supported in part by the U.S.~Department of Energy under grant No.~DE-SC0007914, and by the PITT PACC.

\section*{Appendix}

\noindent {\bf Width of a three-body charged meson decay.} For a given charged meson $\mathfrak{m}$, the width of its three-body decay $\mathfrak{m}\to \ell \nu_\ell V$ ($\ell = e,~\mu$) is given by
\begin{equation}
    \Gamma(\mathfrak{m}\to \ell \nu_\ell V) =\frac{1}{64\pi^3 m_{\mathfrak{m}}}\int_{E_\ell^-}^{E_\ell^+} dE_\ell \int_{E_\nu^-}^{E_\nu^+}dE_\nu \overline{ \left| \mathcal{M} \right|^2},
\end{equation}
for which the integration ranges are
\begin{eqnarray}
    E_\ell^-&=&m_\ell,~~E_\ell^+=\frac{m_\mathfrak{m}^2+m_\ell^2-m_V^2}{2m_{\mathfrak{m}}}, \\
    E_\nu^{\pm} &=& \frac{m_{\mathfrak{m}}^2+m_\ell^2-m_V^2-2m_{\mathfrak{m}}E_\ell}{2(m_{\mathfrak{m}}-E_\ell \mp p_\ell)}.
\end{eqnarray}

\medskip

\noindent {\it Case (i)}: $\nu$-philic model. The mediator $V$ can be emitted only from the neutrino leg at tree level as shown in Fig.~\ref{fig:diagrams}($b$) and the resulting matrix element can be written as 
\begin{eqnarray}
    i\mathcal{M}&=&\frac{G_F}{\sqrt{2}}g_V V_{\mathrm CKM}^{\mathfrak m} f_\mathfrak{m} \varepsilon^\mu\left[\bar{u}_\ell p_{\mathfrak m}^\rho \gamma_\rho \frac{\cancel{p}_V+\cancel{p}_\nu}{(p_V+p_\nu)^2} \right. \nonumber \\
    &\times& \left.\gamma_\mu(1-\gamma_5)v_{\nu}\right]\,,
\end{eqnarray}
where $\varepsilon^\mu$ in the polarization vector for the outgoing $V, $ $G_F$ is the usual Fermi constant, $f_{\mathfrak m}$ is the decay constant of charged meson species $\mathfrak m$, $V_{\rm CKM}^{\mathfrak m}$ is $V_{ud}$ and $V_{us}$ for $\mathfrak m=\pi^\pm$ and $\mathfrak m = K^\pm$, respectively, and $p_i$ denotes the four-momentum of particle species $i$. 
We then find the spin-averaged matrix element squared to be
\begin{widetext}
\begin{eqnarray}
    \overline{\left | \mathcal{M}\right |^2} 
&=& \frac{2 (G_Ff_{\mathfrak m}m_{\mathfrak m}g_V V_{\rm CKM}^{\mathfrak m})^2}{m_V^2(m_{\mathfrak m}^2+m_\ell^2-2m_{\mathfrak m}E_\ell)^2} \left[ 8 E_\ell^3 m_{\mathfrak m}\left\{m_{\mathfrak m}(m_{\mathfrak m}-2E_\nu)-2m_V^2 \right\} \right. \nonumber \\
&+&4E_\ell^2 \left\{2m_V^4 -(7m_{\mathfrak m}^2-4E_\nu m_{\mathfrak m} +2m_\ell^2)m_V^2 +m_{\mathfrak m}(3m_{\mathfrak m}^3-6E_\nu m_{\mathfrak m}^2+3m_\ell^2 m_{\mathfrak m}-4E_\nu m_\ell^2) \right\} \nonumber \\
&-& 2E_\ell \left\{ 4m_{\mathfrak m} m_V^4 +4(m_\ell^2+2m_{\mathfrak m}^2)(E_\nu-m_{\mathfrak m})m_V^2 + (m_\ell^2+m_{\mathfrak m}^2)(3m_{\mathfrak m}^3-6E_\nu m_{\mathfrak m}^2+3m_\ell^2 m_{\mathfrak m} -2E_\nu m_\ell^2)\right\}  \nonumber \\
&+& \left. 2(m_{\mathfrak m}^2-m_\ell^2)m_V^4-(m_{\mathfrak m}^2+m_\ell^2)(3m_{\mathfrak m}^2-4E_\nu m_{\mathfrak m}-m_\ell^2)m_V^2 +(m_{\mathfrak m}^2+m_\ell^2)^2(m_{\mathfrak m}^2-2E_\nu m_{\mathfrak m}+m_\ell^2) \right].
\end{eqnarray}
\end{widetext}

\medskip

\begin{figure}[t]
    \centering
    \includegraphics[width=0.3\textwidth]{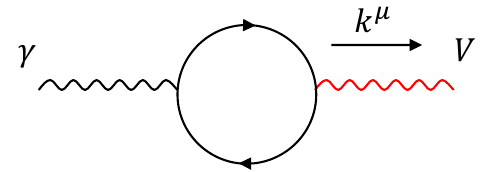}
    \caption{Mixing between the new gauge boson $V$ and the ordinary SM photon via a SM fermion loop.}
    \label{fig:mixing}
\end{figure}

\noindent {\it Case (ii)}: $B-L$ model. In this case, it is instructive to understand first the decay structure relevant to models of $B-3L_\tau$, a variation of the $B-L$ scenario. Since none of the initial and final state particles in the decay process are charged under the $B-3L_\tau$ massive vector field in this model, the vector boson emission occurs through kinetic mixing with the ordinary SM photon. As $\mathfrak m$ and $\ell$ are electrically charged, Figs.~\ref{fig:diagrams}($a,c,d$) are relevant. 
The kinetic mixing parameter is given by
\begin{equation}
    \epsilon_V = \sum_f \frac{2\alpha}{\pi} \int_0^1 dx (x-1)x\log\left( \frac{\Lambda^2}{m_f^2+x(x-1)k^2}\right),
\end{equation}
where $f$ runs over all fermion degrees of freedom charged under $V$ and the SM photon field, $\alpha$ is the usual fine structure constant, and $k$ is the momentum of the photon and $V$ propagators (see Fig.~\ref{fig:mixing}).
We evaluate $\epsilon_V$ at $k^2 = m_V^2$ and find that the dependence of $\epsilon_V$ on $m_V^2$ and $\Lambda^2$ is mild. We choose a fixed value of $\epsilon_V$ over the relevant $m_V$ values for illustration purposes:
\begin{equation}
    \epsilon_V \approx -0.01.
\end{equation}

Taking the kinetic mixing parameter into account, we find that the most general matrix element for the decay process at hand can be written as 
\begin{eqnarray}
    i\mathcal{M}&=& - \frac{G_F}{\sqrt{2}}\epsilon_{V} g_V V_{\rm CKM}^{\mathfrak m} \varepsilon_\mu \left[ \bar{u}_\ell \gamma_\rho (1-\gamma_5)v_\nu  T^{\mu\rho} \right.\nonumber \\
    &-& \left. f_{\mathfrak m} \bar{u}_\ell \gamma^\mu \frac{\cancel{p}_\ell+\cancel{p}_V+m_\ell}{(p_\ell+p_V)^2-m_\ell^2} p_{\mathfrak m}^\rho\gamma_\rho(1-\gamma_5)v_\nu \right],
\end{eqnarray}
where $T^{\mu\rho}$ is a hadronic tensor including QCD structure-dependent form factors. 
The most generic form of $T^{\mu\rho}$ is given by~\cite{Khodjamirian:2001ga}
\begin{eqnarray}
    T^{\mu\rho}&=&c_1 g^{\mu\rho}+c_2(p_\ell+p_\nu)^\mu p_V^\rho+c_3(p_\ell+p_\nu)^\rho p_V^\mu \nonumber \\
    &+& c_4(p_\ell+p_\nu)^\mu (p_\ell+p_\nu)^\rho+c_5 p_V^\mu p_V^\rho \nonumber \\ &+& F_V\epsilon^{\mu\rho\lambda\sigma}(p_\ell+p_\nu)_\lambda p_{V,\sigma}\,,
\end{eqnarray}
where $c_i$ are form factor parameters or coefficients and $F_V$ is the hadronic form factor of the vector current. 

The squared matrix element is not very illuminating, but one can find that the terms proportional to $c_3$ and $c_5$ eventually vanish and we end up with $c_1$, $c_2$, $c_4$, and $F_V$. 
In principle, these dimensionful parameter values (together with coupling $g_V$ that appears as an overall factor) would be determined by experiments. 
While they are unknown as of now, one could attempt to estimate the scales of these parameters. For example, if $V$ {\it were} massless, the Ward identity or the current conservation holds, resulting in the following relations~\cite{Khodjamirian:2001ga}: 
\begin{eqnarray}
    c_1+c_2(p_\ell+p_\nu)\cdot p_V &=& f_{\mathfrak m}, \\
    c_4(p_\ell+p_\nu)\cdot p_V &=& f_{\mathfrak m}.
\end{eqnarray}
A massive $V$ does not necessarily obey the above relations, but one could still assume $c_1 \sim f_{\mathfrak m}$ and $c_2 \sim c_4 \sim \frac{f_{\mathfrak m}}{(p_\ell+p_\nu)\cdot p_V}$. In this context, we take the following reference parameter values (form factor parameter Choice I):
\begin{equation}
    c_1 = 0.1~{\rm GeV},~(c_2,c_4,F_V)=(10,10,0.2)~{\rm GeV}^{-1}\,, \label{eq:params}
\end{equation}
where the $F_V$ value is inferred from the $\pi^+ \to e^+\nu_e\gamma$ consideration~\cite{Donoghue:1992dd}. The form factor parameter values for $K^\pm$ are not necessarily the same as those for $\pi^\pm$, but we set both sets of parameter values to be the same for purposes of illustration.\footnote{We do not use the photon form factors $F_V$ and $F_A$ for the dark photon case as in Ref.~\cite{Chiang:2016cyf}.} 

When it comes to the $B-L$ gauge boson case, it can directly couple to the charged lepton and the neutrino, whereas it can interact with the charged meson again via kinetic mixing. 
The resulting matrix element takes the form of
\begin{eqnarray}
    i\mathcal{M}&=& - \frac{G_F}{\sqrt{2}} g_V V_{\rm CKM}^{\mathfrak m} \varepsilon_\mu \left[ \epsilon_{V}\bar{u}_\ell \gamma_\rho (1-\gamma_5)v_\nu  T^{\mu\rho} \right.\nonumber \\
    &-&  f_{\mathfrak m} \bar{u}_\ell \gamma^\mu \frac{\cancel{p}_\ell+\cancel{p}_V+m_\ell}{(p_\ell+p_V)^2-m_\ell^2} p_{\mathfrak m}^\rho\gamma_\rho(1-\gamma_5)v_\nu \nonumber \\ 
    &+& \left. f_{\mathfrak m} \bar{u}_\ell p_{\mathfrak m}^\rho\gamma_\rho \frac{\cancel{p}_\nu+\cancel{p}_V}{(p_\nu+p_V)^2} \gamma^\mu(1-\gamma_5)v_\nu \right] .
\end{eqnarray}
Note that the third term has a sign opposite to the second term because the charged lepton (anti-charged lepton) comes along with an antineutrino (neutrino), and therefore, the two terms interfere destructively. 

The squared matrix element is not illustrative either, so we again omit the expression here. When evaluating the partial decay widths, we take the same form factor parameter values as in Eq.~\eqref{eq:params}. To illustrate the impact of different parameter choices, we consider a different set of parameter values (form factor parameter Choice II) as follows:
\begin{equation}
    c_1 = 10^3~{\rm GeV},~(c_2,c_4,F_V)=(10^4,10^4,0.2)~{\rm GeV}^{-1}\, . \label{eq:params2}
\end{equation}
The two choices~\eqref{eq:params} and \eqref{eq:params2} are used to obtain the numerical results shown in Figs.~\ref{fig:B-3Ltau}($b,c$), respectively. 

\medskip

\noindent {\bf Modeling of the focusing-horn effects.} DUNE ND-LAr is an on-axis detector and the expected neutrino flux is affected by the magnetic horn configuration. Whether the produced charged mesons are focused for ND-LAr depends on their production angle and energy. In our simulation, we assume that if the energy $E$ and angle $\theta$ of a given charged meson are within certain ranges, its momentum direction gets perfectly aligned to the beam axis and decays before reaching the shielding area according to its associated decay law.

We attempt to reproduce the differential $\nu_\mu$ fluxes in Ref.~\cite{DUNE:2020ypp}, which we denote by $d\Phi_\nu^{\rm DUNE}/dE_\nu$, using this simplified simulation scheme. We find that the following combination of two differential neutrino flux components $d\Phi_\nu^1/dE_\nu$ and $d\Phi_\nu^2/dE_\nu$ can reproduce $d\Phi_\nu^{\rm DUNE}/dE_\nu$ fairly well:
    \begin{equation}
        \frac{d\Phi_\nu^{\rm DUNE}}{dE_\nu} \approx  3 \frac{d\Phi_\nu^1}{dE_\nu} + 0.2 \frac{d\Phi_\nu^2}{dE_\nu}\,,
    \end{equation}
    where the prefactors for $d\Phi_\nu^{1,2}/dE_\nu$ are rescaling factors and $\Phi_\nu^{1,2}$ consist of the neutrinos from the charged mesons defined by
    \begin{eqnarray}
        \Phi_\nu^{1}&:& E_{\pi/K}<10~{\rm GeV},~\theta_{\pi/K} \in [0.01, 1]~{\rm rad}, \\
        \Phi_\nu^{2}&:& E_{\pi/K} <120~{\rm GeV},~\theta_{\pi/K} \in [0.01, 1]~{\rm rad}.
    \end{eqnarray}

In addition, we consider the $\nu_\tau$-optimized mode discussed in Ref.~\cite{DUNE:2020ypp}, where the $\nu_\mu$ fluxes expected at the far detectors (FD) in the standard and $\nu_\tau$-optimized modes are reported. We observe that the standard-mode $\nu_\mu$ fluxes at both ND and FD are very similar to each other up to a normalization factor. Our simulated differential $\nu_\mu$ flux defined by $E_{\pi/K}\in [5,~120]~{\rm GeV},~\theta_{\pi/K} \in [0.01,~1]~{\rm rad}$ reproduces the corresponding flux fairly well when normalized by a rescaling factor of 1.5.  

 As for ICARUS-NuMI, since the detector is off the beam axis, the neutrino flux at the detector benefits from the magnetic horn in a limited manner. It turns out that the assumption of isotropic charged-meson fluxes can describe the neutrino fluxes reasonably well~\cite{private}. 
We find that $\nu_\mu$ spectra can be described by the $\nu_\mu$'s from isotropic charged mesons together with an overall rescaling factor of 30.

\bibliography{ref}

\end{document}